\documentclass[english,prd,onecolumn,nofootinbib]{revtex4}
\pdfoutput=1
\usepackage[final]{graphicx}
\usepackage{bm, color}
\usepackage{hyperref}
\usepackage{amsmath,mathrsfs,bbm}
\usepackage{wrapfig}
\usepackage[leftcaption]{sidecap}
\usepackage{color}

\def\be{\begin{equation}}
\def\ee{\end{equation}}
\def\bea{\begin{eqnarray}}
\def\eea{\end{eqnarray}}

\def\d {\mathrm{d}}

\newcommand{\e}{{\bm{{p}}}}
\newcommand{\pe}{{{{p}}}}

\newcommand{\J}{{\bm{\mathcal{J}}}}
\newcommand{\K}{{\bm{\mathcal{K}}}}

\newcommand{\C}[3]{{{\mathscr{C}}^{#2(#1)}_{#3}}}

\newcommand{\R}{{\bm{\mathcal{R}}}}

\begin{document}

\title{Roulettes: A weak lensing formalism for strong lensing  \\--- I. Overview ---}
\author{Chris Clarkson}
\address{Department of Mathematics \& Applied Mathematics, University of Cape Town, Cape Town 7701, South Africa.\\ 
School of Physics \& Astronomy, Queen Mary University of London, Mile End Road, London E1 4NS, UK.}
\email{chris.clarkson@gmail.com}
\date{\today}

\begin{abstract}

We present a new perspective on gravitational lensing. We describe a new extension of the  weak lensing formalism capable of describing strongly lensed images. By integrating the non-linear geodesic deviation equation, the amplification matrix of weak lensing is generalised to a sum over independent amplification tensors of increasing rank.
We show how an image distorted by a generic lens may be constructed as a sum over `roulettes', which are the natural curves associated with the independent spin modes of the amplification tensors. Highly distorted  images can be constructed even for large sources observed near or within the Einstein radius of a lens where the shear and convergence are large. 
The amplitude of each roulette is formed from a sum over appropriate derivatives of the lensing potential. Consequently, measuring these individual roulettes for images around a lens gives a new way to   reconstruct a strong lens mass distribution without requiring a lens model.  This formalism generalises the convergence, shear and flexion of weak lensing to arbitrary order, and provides a unified bridge between the strong and weak lensing regimes.

  This overview paper is accompanied by a much more detailed Paper II, \href{http://arxiv.org/abs/1603.04652}{arXiv:1603.04652}.

\end{abstract}

\maketitle

{\bf \emph{Introduction}}\\

Gravitational lensing is typically though of in two separate regimes -- strong and weak. The theory of lensing is formulated from integrating the geodesic deviation equation and forming the Jacobi map  (see e.g.,~\cite{Bartelmann:2010fz} for a review). The Jacobi map has a corresponding amplification matrix which contains the familiar shear and convergence. In the weak lensing approximation, these are considered to be constants over a small image, and the Born approximation is used whereby the light travels along a straight line. Weakly lensed arclets can be described by including flexion,  which to a good approximation is the derivative of the amplification matrix at the centre of the image~\cite{Goldberg:2004hh,Bacon:2005qr}, and accurately given by solving the  the geodesic deviation at second-order in the deviation vector~\cite{Clarkson:2015pia}. Strong lensing situations, on the other hand, can be analysed by tracing families of geodesics around a given mass distribution, or by analysing how the Jacobi map varies over a lens distribution~\cite{Kochanek:2004ua}. This suffers from the drawback of requiring a fully 3D lens model, whereas for weak lensing we only require the lensing potential at the centre of the image.   

Here we present a new approach to the strong lensing regime, which keeps with the philosophy of the weak lensing approach. The idea is to build up a family of higher-order maps, generalising the linear Jacobi, and second-order Hessian maps~\cite{Clarkson:2015pia}, from which a complicated image can be formed by summing over these (constant) maps. We therefore generalise the weak lensing convergence, shear and flexion expansion to arbitrary order. This means that an image can be constructed as a sum over curves of different spins we  refer to as roulettes, as they are formed from curves generated by circles rolling on circles. In the case of a weak gravitational field, the amplitudes of these are set by the lensing potential and its derivatives at the centre of the image. More generally they are governed by  derivates of the optical tidal tensor integrated along the line of sight to the centre of the image. Consequently, we do not need to know the lens distribution in total to form a large strongly lensed image, just the potential at the centre of the image. Conversely, given a lensed image with good enough resolution one can in principle construct the projected lens mass and its derivatives at the position of the image. This gives a  new way to construct the lens mass distribution without having to assume a model. 

This paper is a highly condensed presentation of Paper II~\cite{Clarkson:2016ccm}, and we refer there for derivations and further details.

~\\
{\bf \emph{Geodesic deviation and weak lensing theory}}\\

The starting point for the computation of the convergence and shear is  from the geodesic deviation equation which is linear in the deviation vector $\xi^a$.\footnote{$a,b,c,\cdots$ denote spacetime indices, $A,B,C,\cdots$ are tetrad indices in the screen space. $R^a_{~bcd}$ is the Riemann tensor. A dot is a derivative along the null curve with tangent vector $k^a$ pointing from source to observer, with the convention $\dot{}=k^a\nabla_a=-\d/\d\chi$. Bold symbols denote vectors and rank-2 tensors in the screen space. See Paper II and~\cite{Clarkson:2015pia} for full details.}
In the screen space this is the key equation for gravitational lensing: 
\be\label{asdjkabfowpjee}
\ddot\xi^A - \mathcal{R}^A_{~~B}\xi^B=0\,, 
\ee
where $\mathcal{R}_{AB}=-R_{AcBd}k^ck^d$ is the optical tidal tensor. In terms of a linear perturbation of flat space,\footnote{We  write our perturbations in the weak field regime with respect to the Poisson gauge, where 
$\d  s^2 =
 \big[-(1 + 2\Phi )\d \eta^2 
 + (1-2 \Psi)\gamma_{i j}\d x^{i}\d x^{j}\big]\,\,.$
} 
the projected part of the Riemann tensor is, to leading order in derivatives of the potential,
\be
\mathcal{R}_{AB}=\delta_{AB}-\nabla_{A}\nabla_B(\Phi+\Psi)\,.
\ee
The solution to \eqref{asdjkabfowpjee} is usually written in term of the Jacobi map $\J$,
\be\label{sjdncskdnj}
\xi_{(1)}^A=\mathcal{J}^A_{~~B}\zeta^B\,,
\ee
where $\bm\zeta$ is the  angle in the image plane at the observer, meaning that the map is normally defined as the map from observer to source. Solving~\eqref{asdjkabfowpjee} to find the perturbed part of the Jacobi map gives:
\be
\mathcal{J}_{AB}=
\chi\mathcal{A}_{AB}=\chi\delta_{AB}-{\chi}^{2}\nabla_A\nabla_{B}\psi\,,
\ee
where we have identified the conventional amplification matrix $\bm{\mathcal{A}}$ in the middle. We use $\chi$ as the radial distance, and the integral is along the past null cone from observer to the source. Here,  $\chi\nabla_A$ is an angular derivative on the observers celestial sphere. The lensing potential is
\be
\psi=\int_0^\chi\d\chi'\left(\frac{\chi-\chi'}{\chi\chi'}\right)(\Phi+\Psi)\,.
\ee 
The trace and trace-free parts of the amplification matrix give rise to the usual convergence ($\kappa$) and shear ($\gamma_{AB}$). In the weak lensing regime, it is often assumed that $\kappa$ and $\gamma_{AB}$ are constant over an image, giving rise to a (de-)magnified and elliptically distorted image. 

This derivation is based on the assumption that the deviation vector is infinitesimal, as the GDE is derived under this assumption. However, for large strongly lensed images this assumption is not appropriate, as null geodesics from different parts of a source are, in effect, non-locally distorted. This means that the lensing potential will vary appreciably over an image, and $\bm\xi$ is no longer `small'. Thus, the notion that shear, for example, elliptically distorts an image applies only infinitesimally.  The next order correction beyond shear and convergence is usually termed flexion, and is accurate to $\mathcal{O}(\xi^2)$~\cite{Goldberg:2004hh,Bacon:2005qr}. Flexion can be approximately calculated by Taylor expanding the amplification matrix in the image plane, but the full derivation of flexion is much more complicated~\cite{Clarkson:2015pia}, arising from a generalised GDE known as the Bazanski equation~\cite{1977AnIHP..27..145B,1977AnIHP..27..115B,Vines:2014oba}, which is accurate to $\mathcal{O}(\xi^2)$. Our aim in this work is to generalise this expansion using in weak lensing to all orders~-- and hence make it work for strongly lensed images. 

~\\
{\bf \emph{Non-linear geodesic deviation and higher-order lensing maps}}\\

The fully non-linear geodesic deviation equation is outstandingly complicated, given by the second derivative of Synge's world function, and its series expansion beyond second-order in $\xi$ is equally horrendous~-- see~\cite{Vines:2014oba}. As far as gravitational lensing is concerned, however, the most important contributions arise from terms with the maximum number of screen space derivatives of the metric potential, and therefore, the Riemann tensor, an approximation we use throughout this work. In this approximation, the generalised GDE is relatively straightforward. Normally, when deriving the GDE, we Taylor expand the connection around the tail of the vector $\bm\xi$, and stop at $\mathcal{O}(\partial\Gamma)$. However, this can be extended to higher derivatives of the connection, as this is just the first term in a series expansion. Consequently, 
we can  replace the optical tidal tensor in \eqref{asdjkabfowpjee} by 
\be
\mathcal{R}^A_{~~B}\mapsto \sum_{n=0}^\infty\frac{1}{(n+1)!}
(\xi^C\nabla_{C})^n \mathcal{R}^A_{~~B}\,.
\ee
to give the leading correction to the GDE (this simple argument can be checked comparing Eq. 5.6 in~\cite{Vines:2014oba}). 
Projecting into the screen space, and ignoring any terms which are not screen space derivatives of the optical tidal tensor, we have
\be\label{kdjscndnsck}
\ddot \xi^A - \mathcal{R}^A_{~~B}\xi^B=\sum_{n=1}^\infty\frac{1}{(n+1)!}
\xi^{A_1}\xi^{A_2}\cdots\xi^{A_n}\xi^B
\nabla_{A_1}\nabla_{A_2}\cdots\nabla_{A_n} \mathcal{R}^A_{~~B}\,,
\ee
to leading order in derivatives of the optical tidal tensor in the screen space.  We can solve this using a perturbation expansion of $\xi^A$, by writing
\be
\xi^A=\sum_{m=1}^{\infty}\frac{1}{m!}\xi^A_{(m)}\,,
\ee
where the $m$'th term in the solution obeys (again keeping the maximum number of derivatives of the optical tidal tensor)
\be\label{dsjkbcsjkdbcskcdb}
\ddot\xi^A_{(m)} - \mathcal{R}^A_{~~B}\xi^B_{(m)} =  \xi^{A_1}_{(1)}\xi^{A_2}_{(1)}\cdots\xi^{A_m}_{(1)}\nabla_{A_1}\nabla_{A_2}\cdots\nabla_{A_{m-1}} \mathcal{R}^A_{~~A_m} \,.
\ee
Note that there is no assumption about the spacetime in doing this perturbation expansion.
The initial conditions at the observer are
\be
{\bm\xi}\big|_\text{observer}=0,~~~~\dot{\bm\xi}\big|_\text{observer}=-\bm\zeta\,.
\ee
Here $\bm\zeta$ is the non-perturbative angle in the image plane at the observer. At first order the solution is~\eqref{sjdncskdnj} which gives the physical distance between two rays at the source to first order given the (exact) observed angular position between the two rays, $\bm\zeta$.  
At each order we can solve \eqref{dsjkbcsjkdbcskcdb} to give the $m$'th-order map, for $m\geq2$,
\be\label{dsjcsjdcn}
\xi_A^{(m)} = \mathcal{M}^\searrow_{AB_1\cdots B_m}\zeta^{B_1}\cdots\zeta^{B_m}\,,
\ee
where
\bea\label{djksncskjdnsjkcfvfkbjdfovj}
\mathcal{M}^{{}^\searrow}_{AB_1\cdots B_m}(\chi) &=& 
\int_{0}^\chi\d\chi'\big[\mathcal{J}_A^{~\,D}(\chi)(\mathcal{J}^{-1})_D^{~~E}(\chi')
\mathcal{K}_E^{~\,F}(\chi')-\mathcal{K}_A^{~\,F}(\chi)
\big]\mathcal{J}_{~~F}^{G}(\chi')\nonumber\\&&
\mathcal{J}^{~\,C_1}_{B_1}(\chi')\cdots\mathcal{J}^{~\,C_m}_{B_m}(\chi')
\nabla_{C_1}\cdots\nabla_{C_{m-1}} \mathcal{R}_{GC_m}(\chi')\,.
\eea
Here $\K$ is the reciprocal Jacobi map, defined to have zero derivative at the observer. The method to find this solution from~\eqref{dsjkbcsjkdbcskcdb} is presented in~\cite{Clarkson:2015pia}.
This family of maps are written in the conventional way whereby $\bm\zeta$ is in the image plane (i.e., the \emph{angular} separation between two observed rays) is mapped to $\bm\xi$ in the source plane (the \emph{physical} distance between the same two rays at emission)~-- we follow~\cite{Lasky:2009ca}. For $m=1$ we write ${\mathcal{J}}_{AB}=\mathcal{M}^{{}^\searrow}_{AB}$, noting that the Jacobi map is defined in this way. Adding up the $\bm\xi_{(m)}$'s for an observed deviation angle $\bm\zeta$ will give the physical position of the source element, assuming the maps are known. 

These generalised Jacobi maps are for a generic optical tidal tensor $\bm\R$, suitable for most spacetimes and situations. Specialising to the case of perturbations about flat space, 
the generic image-to-source map is (for $m\geq2$)
\bea\label{opkdvdfv}
\mathcal{M}^{{}^\searrow}_{AB_1\cdots B_m} 
&=& \chi \mathcal{A}_{AB_1\cdots B_m} = -\chi^{m+2}\nabla_A\nabla_{B_1}\cdots\nabla_{B_m}\psi
\eea
The amplification tensors $\mathcal{A}_{AB_1\cdots B_m}$ are the natural dimensionless extensions of the amplification matrix in the regime where  we can linearise in the metric potentials. 

~\\
{\bf \emph{Spin roulettes}}\\

\begin{figure}[htbp]
\begin{center}
\includegraphics[width=\textwidth]{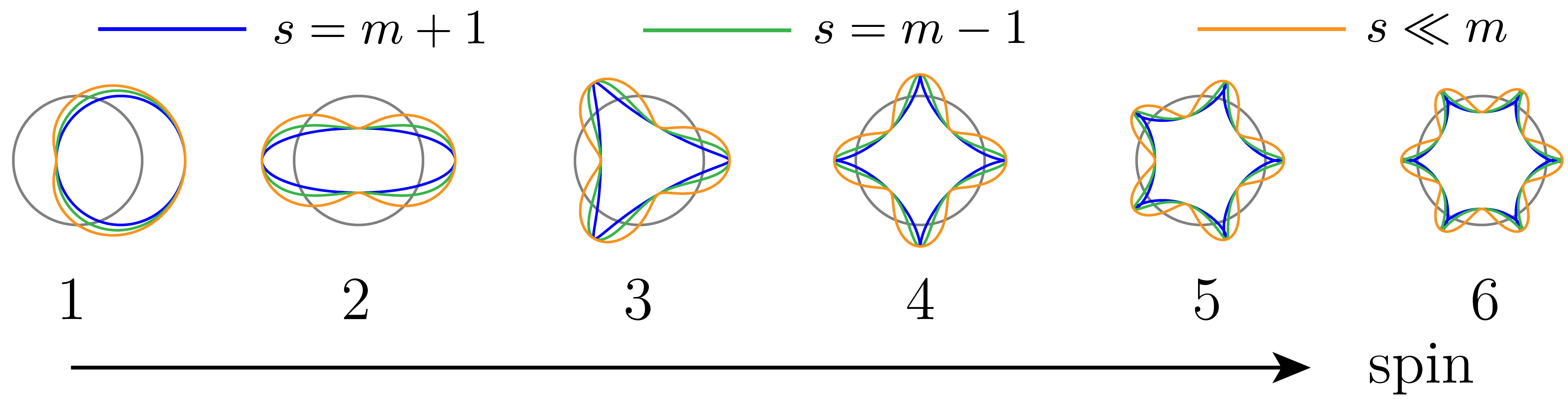}
\caption{Distortions of the grey circle by the $\alpha_s^m$ mode, with the spin shown. Depending on the order of the map, $m$, these spin modes give rise to different roulettes. For example, $\alpha^1_2$ is the blue ellipse which is the normal shear mode, while $\alpha^2_1$ is the green centroid shift of $\mathcal{F}$-type flexion, and $\alpha^2_3$ is the blue triangular spin-3 $\mathcal{G}$-type flexion. All the other curves are new modes which appear at higher order.}
\label{dksnscnksjncskjnc}
\end{center}
\end{figure}

What does a map at each order do? To analyse this, we have to split the map into its normal modes, or spin degrees of freedom. This can be done by separating the tensor into its trace and trace-free parts. Each trace-free tensor then is an independent spin mode (in general, but they are not independent in the weak field regime). The derivation is a bit involved, but the projected map can be invariantly decomposed using coordinates in the screen space such that $\bm\zeta=r\cos\theta\,\bm e_x+r\sin\theta\,\bm e_y$. For the $m$'th map we write the amplification tensor as a sum over independent spin modes of amplitude $\alpha_s^m$ and $\beta_s^m$:
\bea\label{dsifvuheuif}
\mathcal{A}_{AB_1\cdots B_m}\zeta^{B_1}\cdots\zeta^{B_m}
&=&r^m\sum_{s=0}^{m+1} \frac{[1-(-1)^{m+s}]}{4} \nonumber\\&&\hspace{-1cm}\bigg\{
\bigg(1+\frac{s}{m+1}\bigg)\big[\alpha_s^m\bm R_-+\beta_s^m\bm R_/\big]\cdot\e_{(s-1)}
+\bigg(1-\frac{s}{m+1}\bigg)\big[\alpha_s^m\bm I+\beta_s^m\bm \varepsilon\big]\cdot\e_{(s+1)}\bigg\}\,.
\eea
Here we have used the spin vector $\e_{(s)}=\cos s\theta\,\,\bm e_x+\sin s\theta\,\,\bm e_y$ (which traces out a trochoid of spin $s-1$), and the orthogonal matrices 
\be
\bm I          = \left(\begin{array}{cc}1 & 0 \\0 & 1\end{array}\right)\,,~~~
\bm\varepsilon= \left(\begin{array}{cc}0 & 1 \\-1 & 0\end{array}\right)\,,~~~
\bm R_-         = \left(\begin{array}{cc}1 & 0 \\0 & -1\end{array}\right)\,,~~~
\bm R_/          = \left(\begin{array}{cc}0 & 1 \\1 & 0\end{array}\right)\,.
\ee 
When acting on $\e_{(s)}$, the rotation matrices $\bm I$ and $\bm\varepsilon$ lower the spin by 1, and the reflection matrices $\bm R_-$ and 
$\bm R_/$ raise it by 1. Thus each term in the sum in \eqref{dsifvuheuif} has spin $s$. (The same expansion in the helicity basis is given in Paper II.) The action on a unit circle is shown in Fig.~\ref{dksnscnksjncskjnc}. In general there is a family of odd modes alongside these, but these are zero in the weak field approximation, where only the highest derivatives of the potential are kept.

The amplitudes of the roulette modes can be found by inverting \eqref{dsifvuheuif} if we assume they are constants across the image (which is the standard weak lensing approach to convergence shear and flexion). We can do this by integrating in circles around a specific central point $r=0$. Writing $\hat\xi^{(m)}_A=\mathcal{A}_{AB_1\cdots B_m}\zeta^{B_1}\cdots\zeta^{B_m}/r^m=\frac{1}{\chi m!}\frac{\partial^m}{\partial r^m}\xi_A\big|_{r=0}$, we have for $s\geq1$
\be
\alpha_s^m=\frac{1}{2\pi}\int_{-\pi}^{\pi}\mathrm{d}\theta\,\,\hat{\xi}_{(m)}^A\left[ \pe^{(s+1)}_A+
 R^-_{AB}\pe_{(s-1)}^B \right]\,,~~~~
\beta_s^m= \frac{1}{2\pi}\int_{-\pi}^{\pi}\mathrm{d}\theta\,\,\hat{\xi}_{(m)}^A\left[ \varepsilon_{AB}\pe_{(s+1)}^B+
 R^/_{AB}\pe_{(s-1)}^B \right]\,,
\ee
while 
\be
\alpha_0^m=\frac{1}{2\pi}\int_{-\pi}^{\pi}\mathrm{d}\theta\,\,\hat{\xi}_{(m)}^A \pe^{(1)}_A\,,~~~~\beta_0^m=0\,.
\ee
To compare with the standard weak lensing formalism, the convergence is $\kappa=-\alpha^1_0$, the shear components are $\gamma_1=-\alpha_2^1,~~\gamma_2=-\beta_2^1$, and the flexion amplitudes are $\mathcal{F}=-2\mu^2_1/3$ and $\mathcal{G}=\mu_3^2$, where we define the roulette magnitude as $\mu^m_s=\sqrt{(\alpha^m_s)^2+(\beta^m_s)^2}$.

~\\
{\bf\emph{The thin lens approximation}} \\

We can illustrate how these roulettes form a complete image  in the thin lens approximation. We shall use Cartesian $X,Y$ coordinates centred on the lens in the lens plane, at distance $\chi_L$ from the observer, with the distance from the lens $R=\sqrt{X^2+Y^2}\simeq\chi_L\vartheta$, where $\vartheta$ is the angle at the observer from the centre of the lens. (For a circularly symmetric lens we have the lensing potential $\psi=\psi(R)$.)
In this approximation the amplitude of our roulettes become
\bea
\alpha^m_s&=&-\chi_L^{m+1}\sum_{k=0}^m {m\choose k} \bigg[
\C{k}{m}{s} \partial_X
+
\C{k+1}{m}{s}\partial_Y
\bigg]\partial_{X}^{m-k}\partial_Y^k\psi(X,Y)\,.\label{djksnckjdncsk}
\eea
where $\C{k}{m}{s}=\frac{1}{\pi}\int_{-\pi}^\pi\d\theta\sin^{k}\theta\,\cos^{m-k+1}\theta\,\cos s\,\theta$. The expression for $\beta^m_s$ is the same but with $\cos s\,\theta\mapsto\sin s\,\theta$ in the integrals. Here, the derivatives of the potential should be evaluated at the centre of the image position. 

To illustrate the how a strongly lensed image can be constructed, let us consider an idealised circular lens, a singular isothermal sphere  with potential~\cite{Narayan:1996ba}
$\psi={R_E R}/{\chi_L^2}$, where $R_E$ is the Einstein radius of the lens. From this, we calculate the individual roulette amplitudes, and add up the series of terms to a given $m$. For an isothermal sphere, if $m$ is even (odd) all the non-zero spin modes are odd (even). So, for example, maps which have $m$ odd have a spin-0 mode. This series over $s$ and $m$ gives $\bm\xi\sim(x',y')$ as a rather messy formula for the points in the source plane which map to points  $\bm\zeta=(x,y)=r(\cos\theta,\sin\theta)$ in the image plane. 
Consider for example a  source with a Gaussian intensity profile $I=e^{-({x'}^2+{y'}^2)/2\sigma^2}$. 
Given that $x'=x'(r,\theta)$ and $y'=y'(r,\theta)$, the level surfaces of $I$, which are circles in the source plane,  will give the observed curves in the $(x,y)$ image plane.

\begin{figure}[htbp]
\begin{center}
\includegraphics[width=\textwidth]{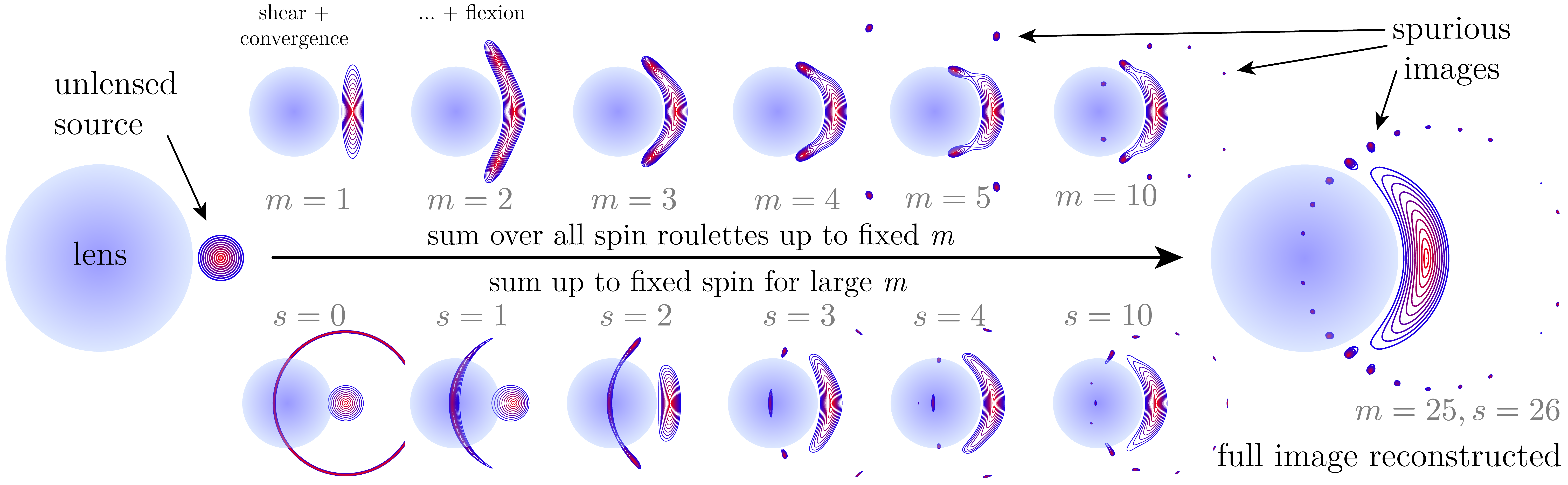}
\caption{A circular isothermal lens, showing how the series of roulettes sums to give a complete primary image when added in different ways. The source is shown on the left and the magnified image on the right (calculated with $m=25$)~-- these are shown to scale, with the Einstein radius of the lens shown for reference. In the top  row we fix the maximum $m$ used, and add up all the possible spins up to $s=m+1$. In the bottom row we add up all possible spins up to the maximum shown for $m\gg s$. Note how poorly the convergence shear and flexion approximate the true image.}
\label{sjakdcdfhbv}
\end{center}
\end{figure}
 We consider in Fig.~\ref{sjakdcdfhbv} how a complete image is formed for such a Gaussian profile. We illustrate this adding up the modes in two different ways. In the top row we add up to a fixed $m$ shown, including all the spin modes up to that order~-- this is the natural way to add up the modes as each mode contributes a factor of $R_E/R$ less. So, for $m=1$, we include the shear and convergence, and for $m=2$ we add in the flexion, and so on. Near the centre of the image this is reasonably accurate, but far from it it is far from it. We also see spurious images appearing in a circle around the centre of the image~-- these appear at the edge of the radius of convergence of the series (note we have removed some of these for clarity). For large $m$ we recover the exact image, with a circle of spurious images. 
 
 However, this is not the only way to order the series. It is also instructive to add in each spin type one at a time~-- we show this in the second row. We fix $m=25$, and add up all the spin-0 modes, then the spin-1 modes and so on, until the same complete image is constructed. The images and circle of spurious images are rather different. The final image shown on the right corresponds to that found by using the exact solution directly. 
 
\begin{figure}[htbp]
\begin{center}
\includegraphics[width=\textwidth]{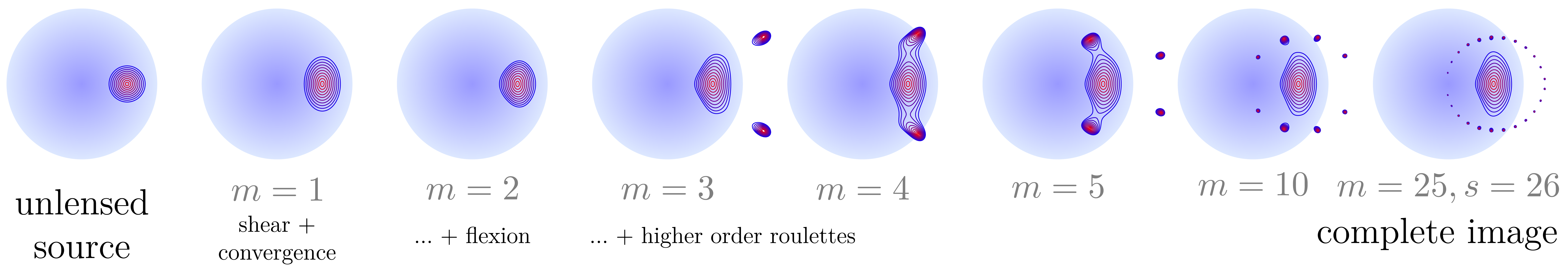}
\caption{A circular lens, showing how the series of roulettes sums to give a complete  image when added up map by map. The source is shown on the left and the magnified image on the right (calculated with $m=25$)~-- these are shown to scale, with the Einstein radius of the lens shown for reference. As in Fig.~\ref{sjakdcdfhbv} we see a ring of spurious images at the radius of convergence. }
\label{sjakdcdfhbvdsccsc}
\end{center}
\end{figure}

 Note that only a single image is reconstructed in this example, with the same location as the `unlensed source' position~-- this is a bit like using the Born approximation. We specify the point where the centre of the image is seen, not where the source is. In general, all images within the radius of convergence of the series will be reconstructed with this prescription. Images within the Einstein radius of the lens can also be accurately made~-- an example is presented in Fig.~\ref{sjakdcdfhbvdsccsc}.

~\\
{\bf\emph{Lens mass reconstruction}}\\

For a circularly symmetric lens the potential can be written in terms of the projected mass 
\be
\psi(R)=\frac{1}{\pi\chi_L^2\Sigma_\text{cr}}\int_0^R\d R_2\frac{M(R_2)}{R_2}\,,
\ee
where the critical surface density is $\Sigma_\text{cr}={2\chi_s}/{\chi_L(\chi_s-\chi_L)}$.
If we can find the roulette amplitudes at a given radius from the centre of the lens, we can find the derivatives of the lens mass at that radius from
\bea
\frac{M}{2\pi R^2\Sigma_\text{cr}} &=& \frac{1}{2}\mu^1_2+\frac{1}{2}\mu^1_0\,,~~~
\frac{M'}{2\pi R\Sigma_\text{cr}} =  \mu^1_0\,,~~~
\frac{M''}{2\pi \Sigma_\text{cr}} = -\frac{2}{3}\left(\frac{R}{\chi_L}\right)\mu^2_1+\mu^1_0\,,\\
\frac{RM'''}{2\pi \Sigma_\text{cr}} &=& \frac{4}{3}\left(\frac{R}{\chi_L}\right)^2\mu^3_0-\frac{2}{3}\left(\frac{R}{\chi_L}\right)\mu^2_1\,,~~~
\frac{R^2M''''}{2\pi \Sigma_\text{cr}} = \frac{8}{3}\left(\frac{R}{\chi_L}\right)^2\mu^3_0+\frac{2}{3}\left(\frac{R}{\chi_L}\right)\mu^2_1-\frac{4}{5}\left(\frac{R}{\chi_L}\right)^3\mu^4_1\,,
\eea
and so on.  This illustrates how, once the amplitudes $\mu^m_s$ are known, the mass and its derivatives can be found. A set of consistency conditions exists between the higher-order roulettes. For a non-symmetric potential, the expressions are more complicated, but the principle remains.

~\\
\noindent{\bf\emph{Conclusions}}\\

We have presented a new formalism for describing large lensed images. The key idea is to extend the normal weak lensing convergence, shear and flexion to all orders so that, in principle, very complicated images can be reconstructed. This results, under the approximations we have made, in a 2D Fourier series expansion over  curves which are all roulettes (epitrochoids plus hypotrochoids) of increasing spin.  Adding up sufficiently many roulettes allows for a complete reconstruction of the image.   In Paper II we show that analytic continuation of the series works for a point mass, and it would be interesting to extend this to more general lenses. The theory should also be developed to include any relevant terms which have been neglected by keeping only the maximal number of screen space derivatives at each order. This was performed for flexion ($m=2$) in~\cite{Clarkson:2015pia}, and is a significant complication. The terms neglected are relatively small, suppressed by factors of $1/\ell^2$ in a harmonic analysis on the observers sphere, but could be important in certain circumstances. As this excites independent `odd' modes discussed in Paper II and~\cite{Clarkson:2015pia}, and alters the consistency relations discussed here, this could provide new tests of modified gravity. 

The amplitude of each roulette depends directly on derivatives of the lensing potential, which offers a new method for the mass reconstruction of a lens.  Consistency conditions between different roulette amplitudes could be used to help identify strong lenses. More generally it will be interesting to see how to construct estimators for the roulette amplitudes, and to see how this technique can be put into useful practise. 
In the weak lensing regime considered in cosmology noise and systematics may make this challenging in the short term, but for moderate to strong lenses simply adding flexion improves cluster substructure reconstruction~\cite{Cain:2015tha}, so it is likely adding more roulettes can improve this further. But the details of this is left for future work.

\acknowledgments 

I would like to thank David Bacon and Julien Larena for discussions and Roy Maartens and Obinna Umeh for comments. This work is funded by the National Research Foundation (South Africa).


\begin{thebibliography}{99}

\bibitem{Bartelmann:2010fz} 
  M.~Bartelmann,
  Class.\ Quant.\ Grav.\  {\bf 27}, 233001 (2010)
  [arXiv:1010.3829 [astro-ph.CO]].


\bibitem{Goldberg:2004hh} 
  D.~M.~Goldberg and D.~J.~Bacon,
  Astrophys.\ J.\  {\bf 619}, 741 (2005)
  [astro-ph/0406376].

\bibitem{Bacon:2005qr} 
  D.~J.~Bacon, D.~M.~Goldberg, B.~T.~P.~Rowe and A.~N.~Taylor,
  Mon.\ Not.\ Roy.\ Astron.\ Soc.\  {\bf 365}, 414 (2006)
  [astro-ph/0504478].


\bibitem{Clarkson:2015pia} 
  C.~Clarkson,
  JCAP {\bf 1509}, no. 09, 033 (2015)
  [arXiv:1503.08660 [gr-qc]].

\bibitem{Kochanek:2004ua} 
  C.~S.~Kochanek,
  astro-ph/0407232.

\bibitem{Clarkson:2016ccm} 
  C.~Clarkson,
  arXiv:1603.04652 [gr-qc].


\bibitem[Bazanski(1977)]{1977AnIHP..27..145B} Bazanski, S.~L.\ 1977, 
Annales de L'Institut Henri Poincare Section Physique Theorique, 27, 145 


\bibitem[Bazanski(1977)]{1977AnIHP..27..115B} Bazanski, S.~L.\ 1977, 
Annales de L'Institut Henri Poincare Section Physique Theorique, 27, 115 


\bibitem{Vines:2014oba} 
  J.~Vines,
  Gen.\ Rel.\ Grav.\  {\bf 47}, no. 5, 59 (2015)
  [arXiv:1407.6992 [gr-qc]].

\bibitem{Narayan:1996ba} 
  R.~Narayan and M.~Bartelmann,
  astro-ph/9606001.

\bibitem{Lasky:2009ca} 
  P.~Lasky and C.~Fluke,
  Mon.\ Not.\ Roy.\ Astron.\ Soc.\  {\bf 396}, 2257 (2009)
  [arXiv:0904.1440 [astro-ph.CO]].

\bibitem{Cain:2015tha} 
  B.~Cain, M.~Bradac and R.~Levinson,
  arXiv:1503.08218 [astro-ph.CO].


\end{thebibliography}
\end{document}